\documentclass[aip,reprint,amsmath,amssymb,superscriptaddress]{revtex4-2}

\usepackage{graphicx}
\usepackage{dcolumn}
\usepackage{bm}
\usepackage{color}

\begin{document}

\title{Time-dependent density-functional study of intermolecular 
Coulombic decay for 2a$_1$ ionized water dimer}

\author{Kedong Wang}
\email{wangkd@htu.edu.cn}
\affiliation {\it School of Physics, Henan Normal University, Xinxiang 453007, People’s Republic of China}
\author{Cody L. Covington}
\email{cody.covington@vanderbilt.edu}
\affiliation{ Department of Chemistry, Austin Peay State University, Clarksville, 
Tennessee 37044, United States}
\author{Kalman Varga}
\email{kalman.varga@vanderbilt.edu}
\affiliation{ Department of Physics and Astronomy, Vanderbilt University, Nashville, 
Tennessee 37235, United States}

\date{\today}

\begin{abstract}  

A real-space, real-time time-dependent density functional theory (RT-TDDFT) 
with Ehrenfest dynamics is used to simulate intermolecular
Coulombic decay (ICD) processes following 
the ionization of an inner-valence electron. The approach has the advantage of 
treating both nuclear and electronic motion simultaneously, allowing for the 
study of electronic excitation, charge transfer, ionization, and nuclear motion. 
Using this approach, we investigate the decay process for the 2a$_1$ 
ionized state of the water dimer.
For the 2a$_1$ vacancy in the proton donor water molecule, 
ICD is observed in our simulations.
In addition, we have identified a novel dynamical process: at the initial stage, 
the proton generally undergoes a back-and-forth motion. 
Subsequently, the system may evolve along two distinct pathways: 
in one, no proton transfer occurs; in the other, the proton departs 
again from its original position and ultimately completes the transfer process.
In contrast, when the vacancy resides in the proton acceptor water molecule, 
no proton transfer occurs, and ICD remains the sole decay channel.

\end{abstract}

\pacs{31.15.Qg}

\maketitle 

\section{Introduction}

The interparticle Coulombic decay  is a fascinating phenomenon that occurs when an 
excited atom, ion, or molecule in close proximity to another neutral or charged 
species transfers its excess energy to a neighbor via Coulombic interactions. This 
energy transfer results in ionization of the neighboring species, and it is mediated 
by the long-range Coulombic force. The ICD has been investigated theoretically and 
experimentally in rare-gas clusters \cite{07Kulef,19Bennett}, hydrogen-bonded 
systems \cite{11Stoy,24Wang}, and liquid phase \cite{24Zhang,13Thur}. 
A recent review covers the fundamental and applied aspects of the 
ICD and related processes \cite{20Jahnke}.

The research activity related to ICD is intensified in the last few
years. A recent study  showed the role of the  interaction between photoexcited 
$\pi$-electron system  of ICD in pyridine monomers
\cite{Barik2022}. Ultrafast ICD was used to
elucidate the influence of N heteroatoms in non-covalent interactions of aromatic
rings \cite{zhou_revealing_2024,Ren2022}. The role of ICD in hydrogen bond
breaking has also been studied \cite{bejoy_dynamics_2023,doi:10.1021/ja200963y,doi:10.1021/acs.jpclett.2c00957}.
Simulated ICD, where the excess energy needed to ionize a neighbor is
mediated by
photons has also been proposed as a possible mechanism
\cite{cederbaum_stimulated_2024}. Another focus of recent studies has
been the ICD in He nanodroplets caused by double excitations 
\cite{bastian_observation_2024} and other mechanisms
\cite{ben_ltaief_spectroscopically_2024,asmussen_dopant_2023}.

Water is an ubiquitous liquid with unique properties arising from hydrogen bonding, 
which has a profound impact on properties of biological molecules. As a result, water has become 
a focal point of both experimental and theoretical research. One area of particular interest is 
the theoretical prediction of novel electronic relaxation processes that occur following the 
inner-valence ionization of water clusters \cite{97PRLCederbaum}. These studies are essential 
for advancing our understanding of fundamental mechanisms in fields such as radiation chemistry, 
atmospheric science, and biological systems. Among water clusters, the water dimer stands out as 
the simplest hydrogen-bonded system, and it has been extensively studied by various research groups. 
The singly and doubly ionized states of the water dimer have been examined using 
ab initio Green's function methods \cite{06IB, 10Stoy}. In particular, the kinetic energy 
spectrum of the electrons emitted during inner-valence ionization by
the ICD mechanism has been explored in depth.
Jahnke et al. \cite{10Jahnke} investigated the decay process of the 2a$_1$ ionized state in water 
dimers using cold-target recoil-ion-momentum spectroscopy. Their findings revealed that ICD occurs 
more rapidly than proton transfer, leading to the fragmentation of the water dimer into 
two H$_2$O$^+$ ions. Notably, they did not observe proton transfer in their experiments.
In a subsequent study, Richter et al. \cite{18Richter} combined electron–electron coincidence 
spectroscopy with theoretical calculations to investigate proton transfer in water dimers. 
Their findings showed the formation of an H$_3$O$^+$ cation and an OH radical, with theoretical 
analysis focusing on the proton-donating water molecule.
They concluded that proton transfer between neighboring water molecules occurs on the same 
timescale as ICD, a conclusion supported by their experimental results. 
Building on these works, Kumar et al. \cite{24Kumar} used the 
Born–Oppenheimer molecular dynamics (BOMD)
method to study the effect of proton transfer on ICD following 2a$_1$
ionization in water dimers.  The equation of motion coupled cluster
singles and doubles methods combined with complex absorbing potential (CAP)
were used to calculate the lifetime of the ionized state. 
They reported that proton transfer occurs exclusively within 14 fs, while electronic decay 
occurs over 36.6 fs for the vacancy in the proton donor water molecule at ground-state 
geometry. Notably, when the 2s vacancy was on the proton acceptor water molecule, 
no proton transfer was observed, and ICD was the sole decay process.
Recently, Wang et al. \cite{24Wang} simulated ICD processes after the ionization of 
a 2a$_1$ electron in water dimers using real-time time-dependent density functional 
theory. Their simulations successfully capture the ICD mechanism, regardless 
of which water monomer was initially ionized; however, proton transfer was not observed 
in their calculations. 

The time-dependent density functional theory \cite{PhysRevLett.52.997,runge_density-functional_1984,
ullrich_time-dependent_2012} has been extensively developed in the
past decades \cite{Lacombe2023,tokatly_time-dependent_2011,doi:10.1021/acs.chemrev.0c00223,
pavanello_subsystem_2013,elliott_universal_2012,reining_excitonic_2002,
mosquera_fragment-based_2013,zhang_linear-response_2020,shepard_simulating_2021}
and it has been applied to simulate many time-dependent systems 
\cite{PhysRevB.54.4484,10.1063/1.5142502,Varga_Driscoll_2011a,
10.1063/5.0057587,PhysRevA.89.023429,Mrudul2020,PhysRevA.91.023422,
PhysRevA.92.053413,PhysRevB.109.245130,PhysRevA.95.052701,PhysRevA.86.043407}.
The predictive power of the RT-TDDFT calculations depends on the quality of 
the exchange and correlation functionals. It has been shown 
\cite{20Shar} that semilocal density functionals 
might underestimate the proton transfer
probability in small water clusters compared to hybrid functionals.
The charge dynamics, at the same time was found to be very similar for
all functionals tested. Chalabala et al. \cite{chalabala_assessment_2018} 
conducted an extensive  study focusing on the dynamics of ionized water 
dimers. They compared the results from two
nonadiabatic approaches: surface hopping and Ehrenfest dynamics. In
surface hopping, the time evolution of the system occurs on individual
diabatic surfaces, with intersurface transitions determined by the
instantaneous nonadiabatic couplings. In contrast, the Ehrenfest
method uses an averaged diabatic surface to evolve the electronic
states. The authors found that both simulation methods predict similar
statistics of trajectories on short time scale. 
We note that these studies on the  role of hybrid functionals
\cite{chalabala_assessment_2018,20Shar} are based
on adiabatic BOMD and calculations of 
ground state energy differences using hybrid functionals 
and they not use of hybrid functionals in nonadiabatic RT-TDDFT calculations.
This is due to the fact that the computational 
cost of hybrid functionals is very large (up to two orders of magnitude) 
compared to local or semilocal functionals \cite{chalabala_assessment_2018}.  
This is especially true for real-space approaches. Hybrid functionals
are more easily implemented using atom centered
orbitals, but atom centered orbitals are not as flexible for electron 
dynamics as real-space representations. An efficient real-space implementation
of hybrid functionals has been recently reported \cite{10.1063/5.0225396}.


In the present work, we use the RT-TDDFT approach to simulate the ICD processes following 
the ionization of a 2a$_1$ electron in water dimers. Our method employs a real-space grid, 
which allows it to adapt to different system sizes without the constraints of basis 
set representations (used in previous studies, e.g. in Ref. \cite{17PRA-TDDFT-proton}). The real space  approach is particularly useful for systems with complex 
electronic structures where localized or extended excitations play a crucial role. 
Additionally, this work aims to investigate proton transfer and its potential role 
in non-radiative decay mechanisms in the water dimer after the ionization of a 
2a$_1$ electron in a water molecule.

\section{Computation Method}
The computations were performed using density-functional
theory (DFT), with the Kohn-Sham (KS) Hamiltonian of the form
\begin{equation}\label{eq1}
H_{KS}(t) = -\frac{\hbar^2}{2m}\nabla\ ^2 + V_{H}[\rho](r,t) +  V_{XC}[\rho](r,t) + V_{ion}(r,t).
\end{equation}
Here $\rho$ is the electron density, which is defined by a sum over all occupied orbitals:
\begin{equation}\label{eq2}
{\rho}(r,t) = \sum_{k=1}^{N_{orbitals}}{|{\psi}_k(r,t)|}^2,
\end{equation}
V$_H$ is the Hartree potential, defined as 
\begin{equation}\label{eq3}
V_{H}(r,t) = \int{ dr'\frac{\rho(r',t)}{|r-{r'}|} }, 
\end{equation}
accounting for the mean electrostatic interactions from electron-electron 
repulsion. The exchange-correlation potential V$_{XC}$ is 
approximated using the generalized gradient approximation (GGA), 
developed by Perdew et al. \cite{92PRB_GGA}.
V$_{ion}$ is the external potential due to the ions, 
represented by employing norm-conserving pseudopotentials 
centered at each ion as given by Troullier and Martins  \cite{91Trou}. 

The time evolution of the electronic wave function was calculated by
solving the time-dependent KS equation
\begin{equation}\label{eq4}
i\hbar\frac{\partial {\psi}_k(r,t)}{\partial t} = H_{KS}\psi_k(r,t),
\end{equation}
and this equation  was solved by time propagation 
\begin{equation}\label{eq5}
\psi_k(r,t+\delta t) \approx exp \ [-\frac{iH_{KS}(t)\delta t}{\hbar}].
\end{equation}
The time-propagator 
was approximated using a fourth-degree Taylor expansion, given as
\begin{equation}\label{eq6}
\psi_k(r,t+\delta t) \approx \sum_{n=0}^4 \frac{1}{n!}(\frac{-iH_{KS}(t)\delta t}{\hbar})^n\psi_k(r,t)
\end{equation}
A time step of $\delta$t = 1 attosecond is used in the calculations. 
A short time step is necessary to ensure that the time-dependent 
Hamiltonian remains nearly commutative at times t and t + $\delta$t 
and the splitting of the total time-evolution operator 
U(0,$t_{final}$) into successively applied propagators remains valid.
It is also required for the stability of the Taylor-time propagation 
used in the calculations. 

In our simulations, the Ehrenfest approach is used to treat the 
nuclear motion. In this approach, the nuclei move classically 
under the influence of time-dependent quantum forces. The forces 
are calculated as the derivatives of the total energy with respect 
to the ionic positions. The corresponding equations of motion have 
the following form:
\begin{equation}\label{eq7}  
M_i\frac{d^2R_i}{dt^2} = \sum_{i\neq j}^{N_{ions}} \frac{Z_iZ_j(R_i-R_j)}{{|R_i-R_j|}^3} 
- \nabla_{R_i}\int V_{ion}(r,R_i){\rho}(r,t)dr
\end{equation}
where M$_i$, Z$_i$ and R$_i$  are the mass, pseudocharge (valence), 
and position of the $i$th ion, respectively. This differential equation 
was time-propagated using the Verlet algorithm at every time step.

In our RT-TDDFT approach the Kohn-Sham orbitals are represented in a real 
space grid. In practice, these discrete points are organized in a 
uniform rectangular grid, and the accuracy of the simulations are 
controlled by adjusting a single parameter: the grid spacing. In our 
simulations, we used a grid spacing of 0.2 {\AA} and placed 140 points 
along each of the x, y, and z axes. The real space grid approach uses
to zero boundary condition at the cell walls.
When  ionization happens and  the zero-boundary condition
leads to an unphysical reflection of the wave function off the walls 
of the simulation cell. To prevent this, we implemented a CAP with the following form, given by 
Manolopoulos \cite{02Mano}:
\begin{equation}\label{eq8}  
-i\omega(x)= -i\frac{\hbar^2}{2m}(\frac{2\pi}{\Delta x})^2f(y)
\end{equation}
where $x_1$ is the start and $x_2$ is the end of the absorbing 
region, $x$ = x$_2$ - x$_1$, c = 2.62 is a numerical constant, 
m is the electron’s mass, and
\begin{equation}\label{eq9}  
f(y)=\frac{4}{c^2}[\frac{1}{(1+y)^2}+\frac{1}{(1-y)^2}-2],\ y=\frac{(x-x_1)}{\Delta x}
\end{equation}
If the molecule is ionized, electron density
will travel to the edge of the simulation box where it is absorbed
by the CAP. 
The total electron number
\begin{equation}\label{eq10}  
N(t) = \int_v {\rho}(r,t)d^3x
\end{equation}
where V is the volume of the simulation box, will 
therefore diverge from the initial electron number N(0). 
We interpret N(0) - N(t) as the total ionization of the molecule.
The RT-TDDFT code has been described in \cite{16Li,16Russ,17Coving,book}. 

In the calculations, the integration of the electron density in a finite
region around an ion or a charged fragments can lead to fractional
charges. The fractional charge can be interpreted in 
multiple ways. One explanation
is that during the simulation, e.g. approximately 1.8 electrons remain
localized within the molecular region. This fractional charge could
either recombine with the ionized electron cloud or dissociate over a
longer simulation period. In the computational model the electron
charge is also absorbed by the CAP. Alternatively, a more practical
interpretation is that the noninteger charge represents an average -
some molecular fragments may retain 2 valence electrons while others
retain only 1, and the 1.8 figure reflects the mean electron count
per fragment.

\begin{figure}
\includegraphics[width=3.4in]{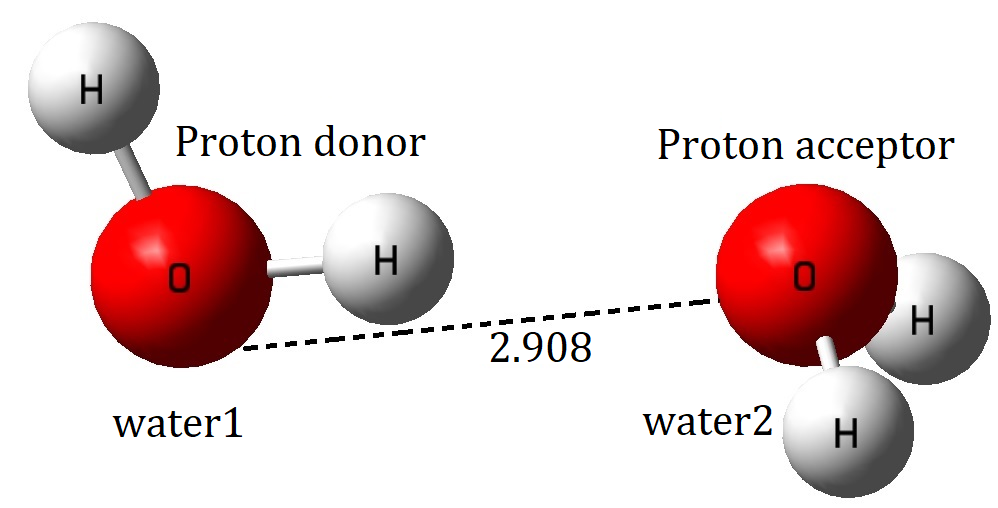}
\caption{\label{dimer} Geometry of the water dimer investigated in 
this work. The O atom internuclear distance is shown in angstrom.}
\end{figure}

Fig. \ref{dimer} presents the geometry of water dimer obtained by
energy minimization using the ground-state DFT. The distance between 
the two O atoms is 2.908 \AA. Our geometry is in good agreement 
with the CCSD(T) optimized geometry \cite{13Lane}. We label water1 
as the hydrogen-bonding proton donor and water2 as the 
hydrogen-bonding proton acceptor in the dimer. As we will discuss 
below, we examine the electron relaxation dynamics following 
the 2a$_1$  inner valence ionization of each molecule in the water dimer.

To initialize the electronic system to account for the initial 
inner-valence ionization, we first perform a ground-state DFT calculation.
After this, we generate the initial hole in
the desired molecular orbital by removing an electron.
We then perform the TDDFT time propagation.

In the calculation presented here, we use a CAP to efficiently 
remove the ionized electron from the system. 
To obtain the charge of the monomer, we integrate the electron density distribution obtained from the DFT calculation with the monomer at the center.

\section{Results}
\begin{figure}[t]
\includegraphics[width=3.4in]{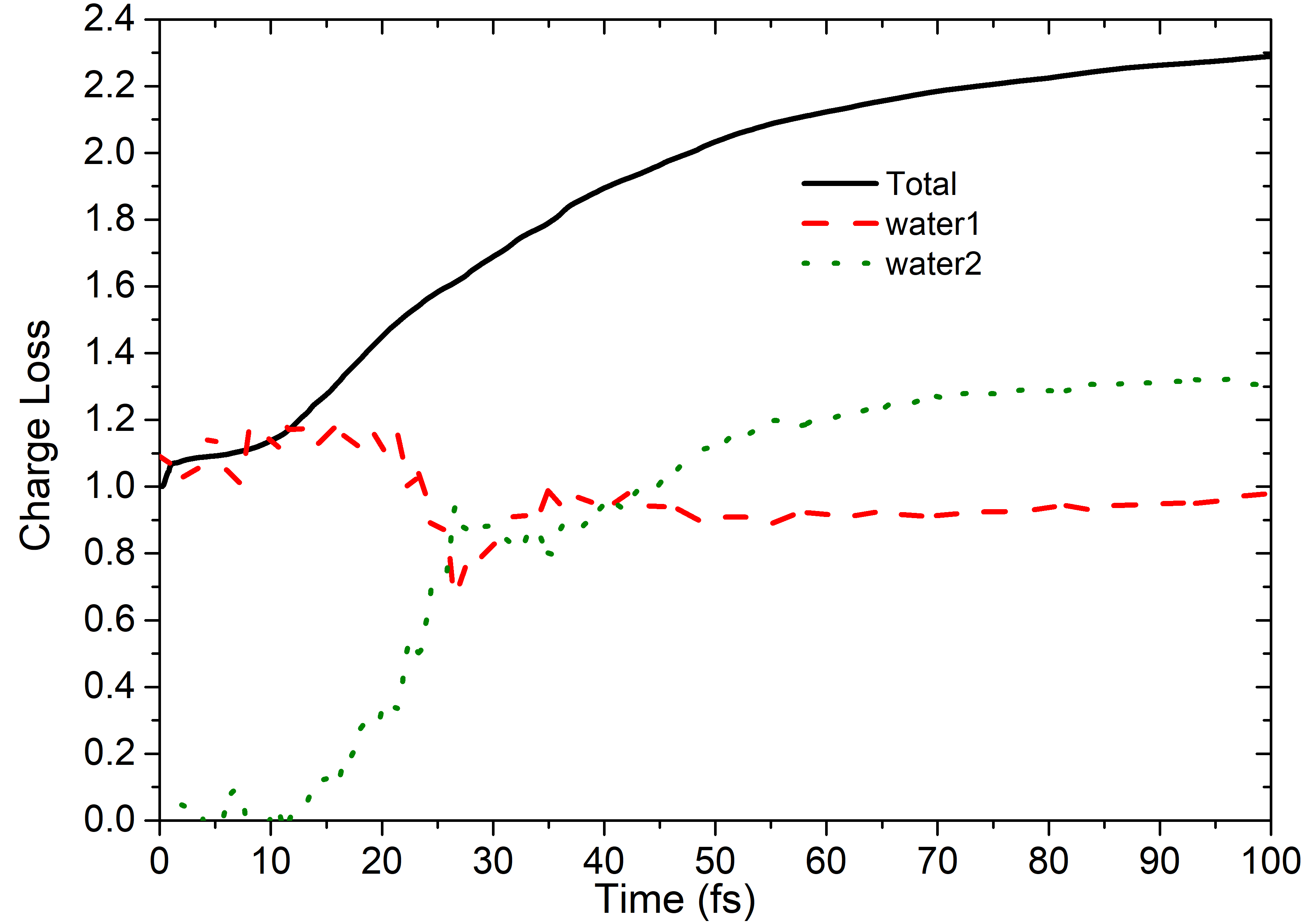}
\caption{\label{water1} Time-dependent charge loss of a water dimer showing proton transfer
when the initial hole is on water1.}
\end{figure}

To explore possible proton transfer events, we conducted multiple dynamical trajectory simulations. When the hole is localized on water1, initial atomic velocities were assigned according to the 
Boltzmann distribution randomizing the initial conditions.
A total of 23 trajectories were simulated (the number of simulations are limited by the
computational cost), among which proton transfer
was observed in 4 cases. 
These four trajectories exhibit qualitatively similar dynamical behavior. As a representative example, 
Fig. \ref{water1} shows the time-dependent charge loss for one of the proton-transferring trajectories.

As shown in Fig. \ref{water1}, the water dimer begins to 
decay following the initial ionization of a 2a$_1$ electron in water1, 
which triggers a secondary ionization on water2. 
The time-dependent charge loss for both water1 and 
water2 is also presented. At the initial time, 
water1 is in a cationic state, while water2 remains neutral.
Throughout the 0–100 fs interval, the charge loss on water1 
remains approximately constant at a value of 1. In contrast, 
the charge loss on water2 begins to increase sharply at around 10 fs, 
reaching a peak of approximately 0.9 by 25 fs. 
Thereafter, from 25 to 100 fs, the charge loss on water2 
fluctuates within the range of 0.9 to 1.3.
These results indicate that the majority of the net charge loss 
occurs on water2, whereas water1 consistently retains its ionic character. 
This behavior is consistent with the characteristics of the ICD mechanism, confirming that our RT-TDDFT approach 
reliably captures the essential features of the ICD process.
The maximum total charge loss is approximately 2.3, 
slightly above the expected value of 2.
However, considering the nature of TDDFT and the dynamics and the influence of the CAP, achieving a perfect final value of 2 is not anticipated, as discussed before. 
The overall charge loss increases slowly, since the CAP is defined in
space, and it takes time for the wavefunction to move to the CAP region.
This makes it difficult to determine the exact time of decay.
A larger box with a CAP that is farther away, would require more time
for the same amount of ionization, but it would require substantially
more computational time. We can still observe a significant increase in the total charge loss between 10-40 fs. Therefore, 
we predict that Coulomb decay occurred on this timescale.

\begin{figure}[t]
\includegraphics[width=3.4in]{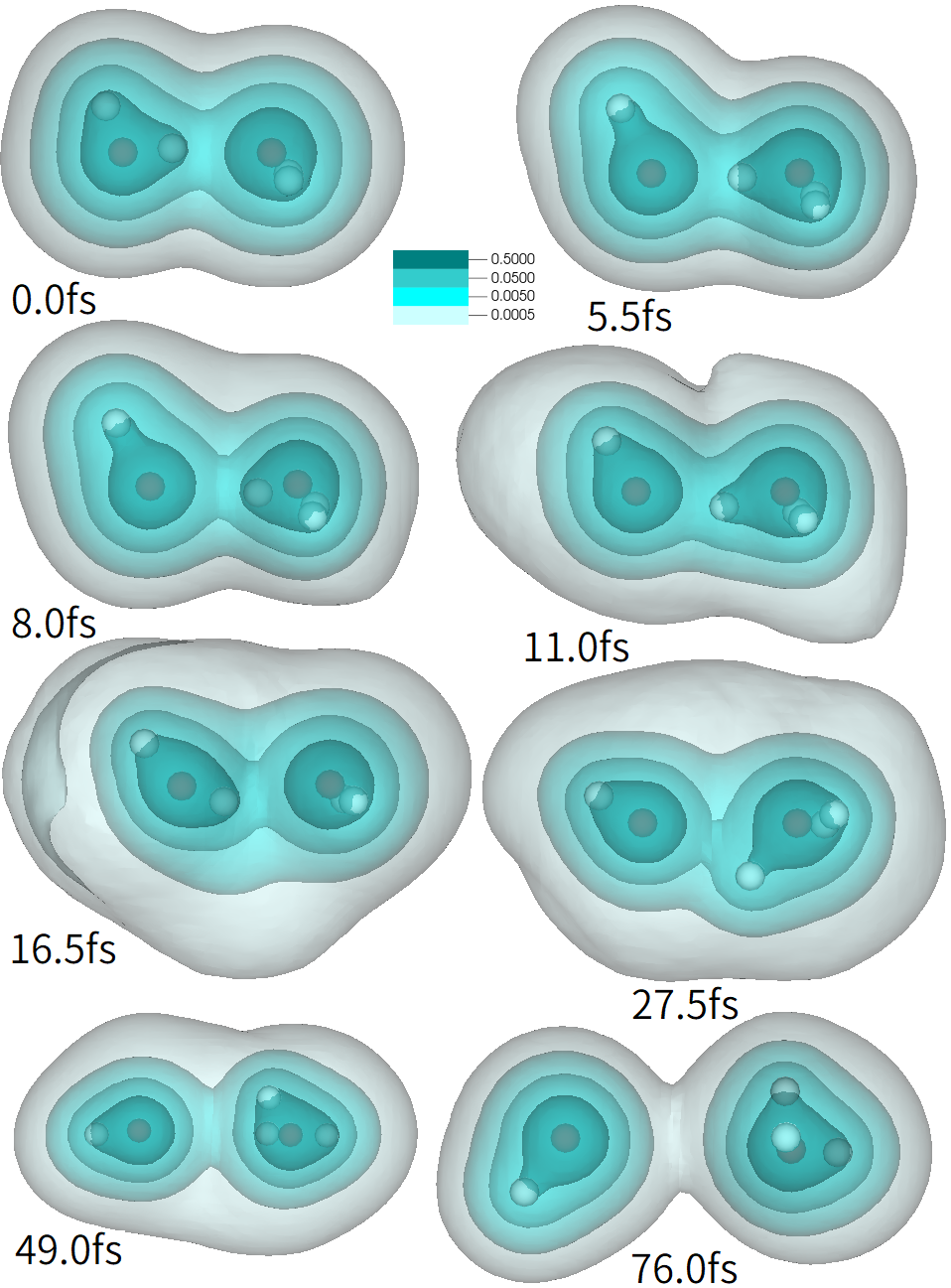}
\caption{\label{TDdensity1} Time-resolved electron density snapshots of a water dimer showing proton transfer when the initial hole is on water1. }
\end{figure}

Further insights into the ICD process can be obtained by examining 
the time-dependent electron density, with selected snapshots presented 
in Fig. \ref{TDdensity1}. A key feature of the dynamics is the 
oscillatory motion of the proton between the two oxygen atoms. 
At 8 fs, the proton is located closest to the oxygen atom in water2. 
It subsequently returns, restoring the original H$_2$O–H$_2$O dimer 
configuration by 16.5 fs. After this point, proton transfer occurs again, 
and the simulation concludes with the formation of H$_3$O$^+$ and OH$^+$ ions. 

\begin{figure}[t]
\includegraphics[width=3.4in]{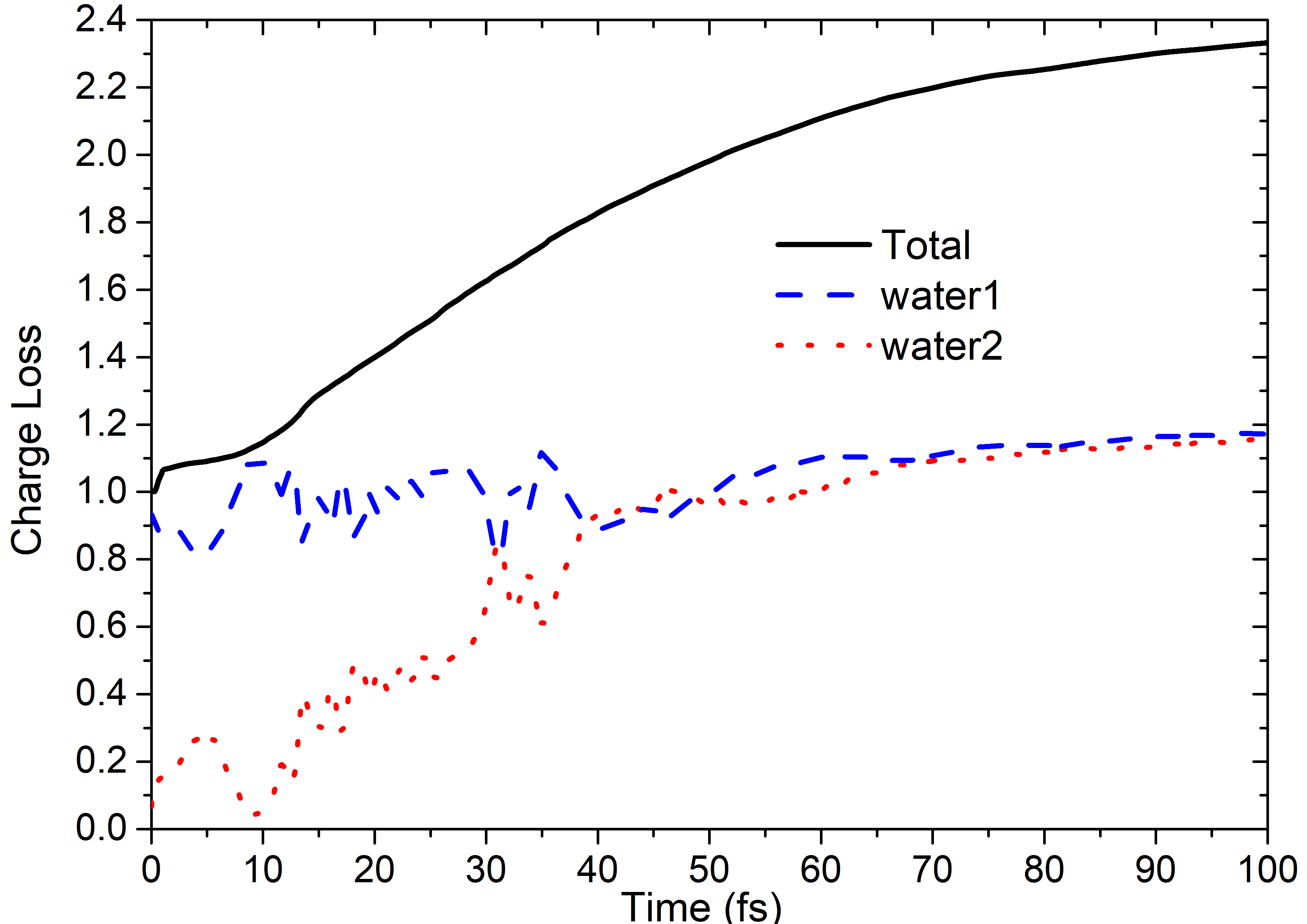}
\caption{\label{water112} Time-dependent charge loss of a water dimer showing no proton transfer
when the initial hole is on water1.}
\end{figure}

\begin{figure}[t]
\includegraphics[width=3.4in]{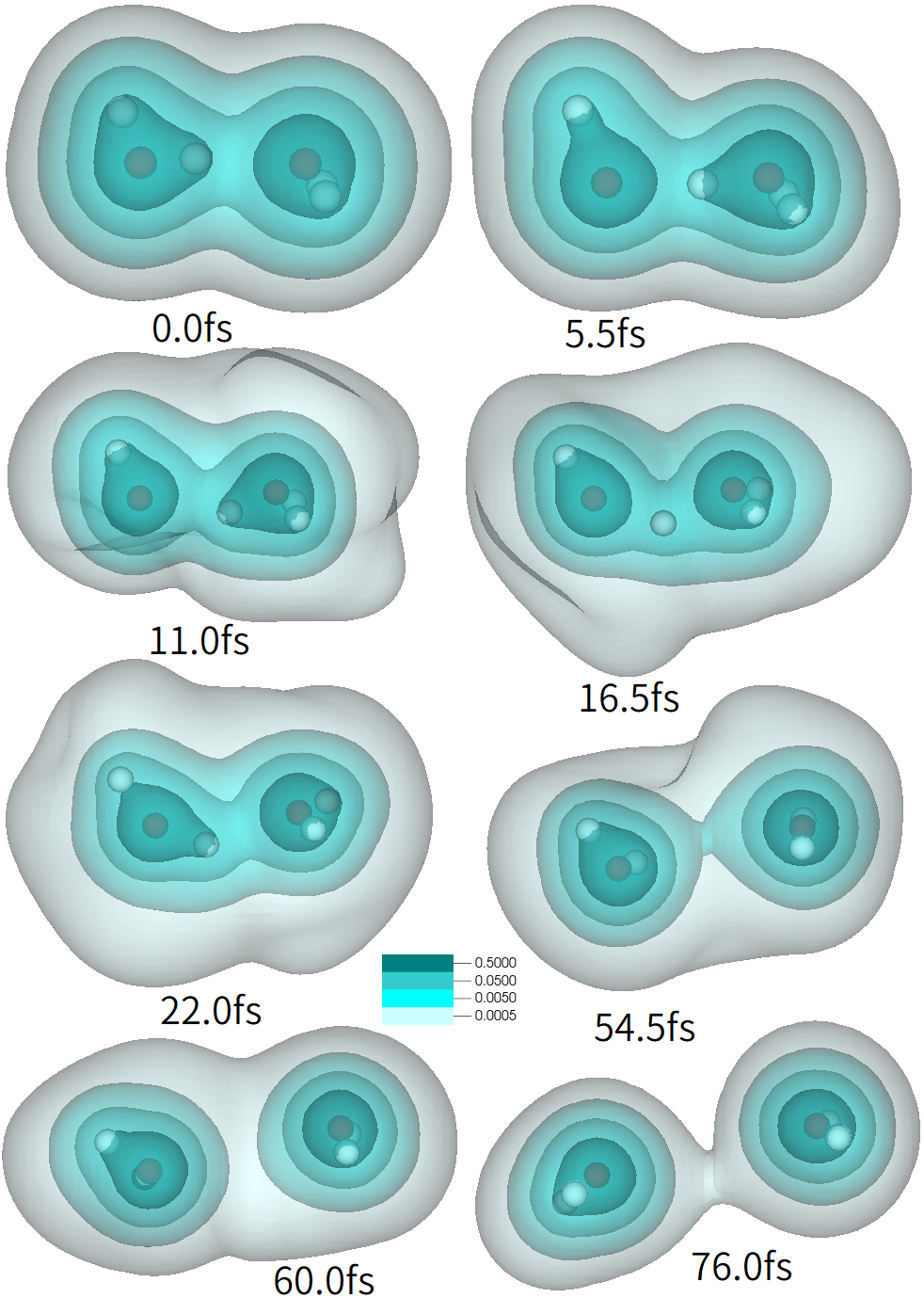}
\caption{\label{TDdensity112} 
Time-resolved electron density snapshots of a water dimer showing no proton transfer when the initial hole is on water1.}
\end{figure}

Now we turn to the other cases where no proton transfer takes place
after a hole is created on water1. The time-dependent charge loss for the dynamics 
is shown in Fig. \ref{water112}. At 0 fs, the total charge loss is approximately 1, originating primarily from water1. 
As the system evolves, water1 remains largely unchanged in its ionic state.
During the initial 0–10 fs period, the total charge loss remains stable 
at around 1. After 10 fs, it begins to increase gradually, 
reaching a maximum value of 2.3 at 100 fs. 
At this point, both water molecules exhibit a charge loss of approximately 1.2, with the predominant contribution arising from water2. 
This charge redistribution is indicative of an ICD process.
The time-resolved electron density snapshots for the dynamics are shown in Fig. \ref{TDdensity112}.
As shown in the figure, during the first 22 fs, the proton undergoes a dynamical process in which it is transferred to water2 and then returns. This back-and-forth motion is similar to the proton transfer observed in Fig. \ref{TDdensity1}. The key difference, however, lies in the fact that here the proton is attracted back by the water1, and no net proton transfer ultimately occurs.
It is noted that this proton transfer motion induces charge redistribution 
between the two monomers, leading to fluctuations 
in their respective charge losses during the 0–20 fs interval, 
as illustrated in Fig. \ref{water112}. 
These observations underscore the coupling between nuclear 
motion and electron dynamics in the ICD process. 

\begin{figure}[t]
\includegraphics[width=3.4in]{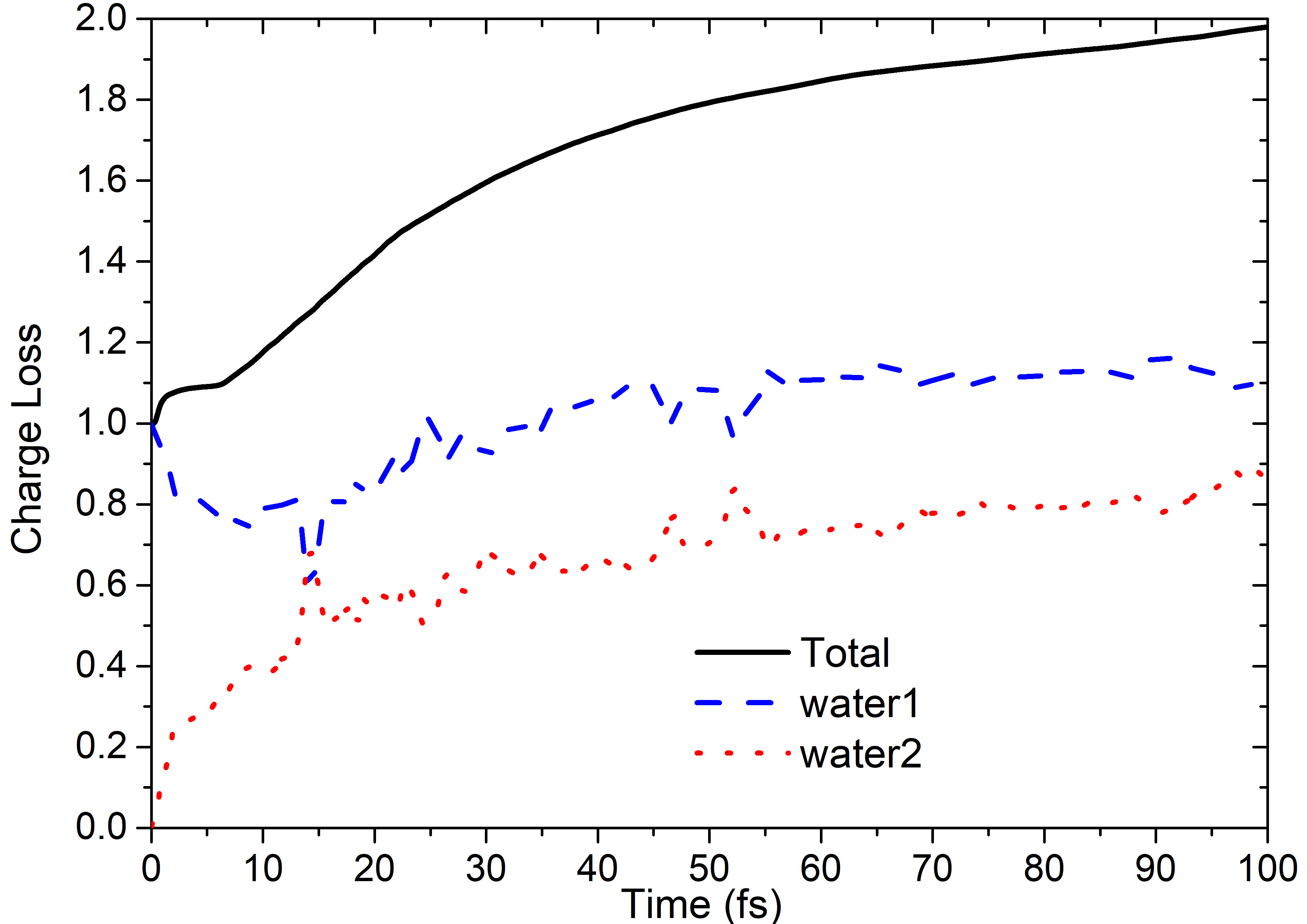}
\caption{\label{water2} Time-dependent charge loss of water dimer when the initial hole is on water2.}
\end{figure}

We have also examined the mechanism of the electronic decay process 
when the vacancy is on the proton acceptor. 
Fig. \ref{water2} shows the time-dependent charge loss 
of water when the initial hole on the water2. 
As shown in the figure, the total charge loss increases gradually from 1.0 at 0 fs to 2.0 at 100 fs.
Starting from 0 fs, the reduction in charge loss of the water1 monomer
occurs rapidly and stabilizes in the range of 0.8 to 1.0 after approximately 15 fs.
Meanwhile, the charge loss of water2 rises steeply within the first 15 fs, followed by a much slower increase, eventually reaching a maximum value of 0.9 at 100 fs.
Considering that charge loss originates primarily from water1, this remains consistent with the ICD mechanism. Due to the slow increase in total charge loss, it is difficult for us to determine the exact time when ICD occurred, but it has indeed occurred within the current 100 fs time frame.

\begin{figure}[t]
\includegraphics[width=3.4in]{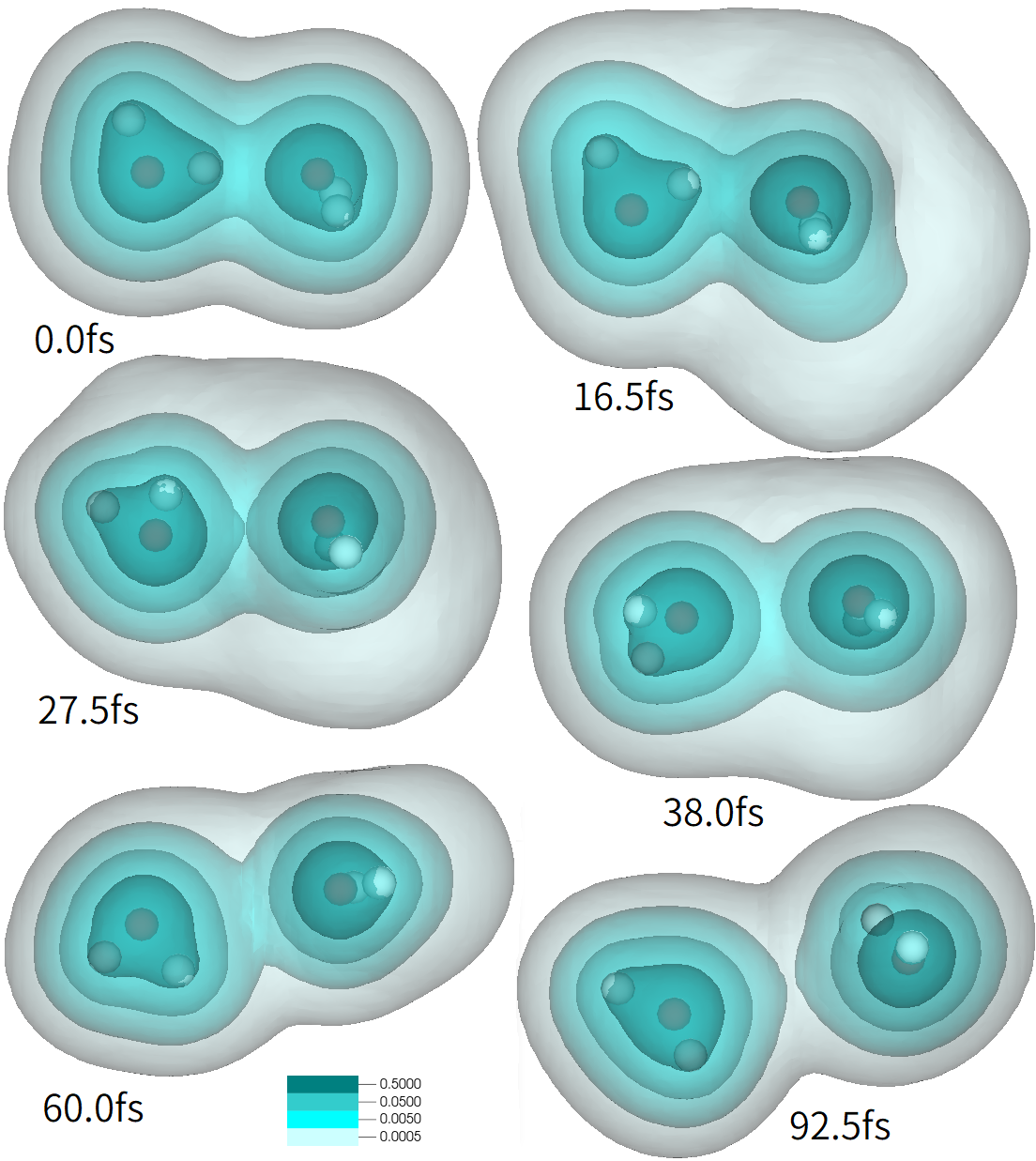}
\caption{\label{TDdensity2} Snapshots of time-dependent electron density of water dimer when the initial hole is on water2. }
\end{figure}

Figure \ref{TDdensity2} presents snapshots of the time-dependent electron density of water dimer when the initial hole is on water2. 
As shown, no proton transfer is observed throughout the simulation time window.
Instead, the water2 undergoes electronic decay, transferring energy to the water1 . This energy transfer induces a rotational motion in the hydrogen atoms: the two hydrogen atoms of water1 rotate approximately 180$^\circ$ around its oxygen atom, while the hydrogen atoms in water2 undergo a similar 180$^\circ$ clockwise rotation around their own oxygen atom.

Finally, we note that while the calculations presented in the paper
are based on an GGA exchange correlation functional, we have also
carried out the same calculations using the local density
approximation (LDA). Our analysis revealed no significant differences
between LDA and GGA functionals in their overall treatment of the
water dimer system. While we did observe some case-by-case
variations—such as individual trajectories showing proton transfer
with LDA but not with GGA under identical conditions
the probability of proton transfer is about the same for both LDA and GGA.

\section{Discussion}

In this section we compare our results with the previous calculations
and experiments. Jahnke et al. \cite{10Jahnke} measured the fragmentation of inner-valence ionized water dimers using cold target recoil-ion-momentum spectroscopy. They observed that ICD occurred faster than proton transfer, leading to the dissociation of the water dimer into two H$_2$O$^+$ ions. In particular, they did not observe decay via proton transfer. 
This experimental observation is consistent with our simulations 
when the vacancy is localized on water2, where ICD is the 
exclusive decay mechanism and no proton transfer occurs.
Furthermore, when the hole is placed on water1, as illustrated in Fig. \ref{TDdensity112}, our simulations indicate that in approximately 80\% of the trajectories, proton transfer does not take place, and ICD remains the sole decay channel. These findings are in good agreement with the results reported by Jahnke et al.

Richter et al. \cite{18Richter} employed electron–electron coincidence spectroscopy to investigate proton transfer in the water dimer, 
observing the formation of an H$_3$O$^+$ cation and an OH radical. 
They also conducted theoretical studies using BOMD and Fano configuration interaction (Fano-CI) methods, focusing on the case where the vacancy is localized on the proton-donor water molecule. Their results indicated that the timescale of proton transfer is comparable to that of ICD, and this was supported by their experimental observations. In our simulations, 
when the vacancy is placed on water1 (the proton donor), 
proton transfer occurs in approximately 20\% of the trajectories. 
As shown in Fig. \ref{TDdensity1}, the proton transfer takes place within the first 30 fs, coinciding with the timescale of ICD. 
This results in the formation of an H$_3$O$^+$ cation 
and an OH radical, in excellent agreement with both the experimental and theoretical findings of Richter et al.

Wang et al. \cite{24Wang} conducted similar calculations to ours using
RT-TDDFT. Their methodology successfully captures the ICD mechanism, regardless 
of which monomer is initially ionized. In their study, both H$_2$O-H$_2$O cases 
exhibit an ICD process occurring on a timescale of 20–30 fs, consistent with our 
conclusions for both scenarios. However, they did not observe proton transfer 
in their simulations following the initial ionization from the 2a$_1$ molecular 
orbital of water1. In contrast to their approach, we employ a real-space grid, 
avoiding the complexities and limitations associated with selecting, converging, 
and managing a basis set. To eliminate any potential differences caused by the 
initial structure of the water dimer, we also tested the initial geometry they 
provided and repeated the calculations, still observing proton transfer.
Recently, Sharma and Fernández-Serra \cite{20Shar} studied the proton transfer 
mechanism after photoionization of H-bonded water molecular chains as a function 
of chain length using nonadiabatic real-time TDDFT simulations in the 
Ehrenfest approximation. They found that the adiabatic scheme (BOMD) at 
comparable timescales failed to describe the evolution of the excited 
system toward a proton-transfer reaction. Thus, the inclusion of
nonadiabatic effects, like in the present work,
was essential for capturing the proton-transfer dynamics.

Kumar et al. \cite{24Kumar} investigated the influence of proton transfer 
on ICD after 2a$_1$ ionization of water 
within a water dimer system using BOMD simulation. 
According to their result, proton transfer occurs exclusively 
when 2a$_1$ ionized state is on water1 with a timescale of 14 fs. 
In contrast, our results reveal a more complex proton dynamics 
under the same ionization condition, 
involving a back-and-forth motion of the proton.
This bidirectional movement is completed within approximately 16 fs.
Following this phase, the system bifurcates into two distinct dynamical pathways. In the first, the proton remains localized near water1, 
and no subsequent transfer occurs, 
resulting in the formation of two water-derived ionic fragments. 
In the second, the proton continues its migration toward water2 
and is ultimately captured, completing the proton transfer 
and yielding an H$_3$O$^+$ cation and an OH radical. 
It is important to emphasize that the structural configuration 
of the water dimer used in our simulations is identical to 
that employed by Kumar et al.\cite{24Kumar}, 
ensuring a consistent basis for comparison.
The reason for the different dynamics is the inclusion of the
nonadiabatic effects.
When 2a$_1$ ionized state is on water2, our results agree with those of Kumar et al.: no proton transfer occurs, and ICD remains the sole decay pathway.

\section{Conclusion}

A real-space and real-time TDDFT approach is used to directly simulate ultrafast electron 
relaxation dynamics following a 2a$_1$ inner-valence electron ionization in water dimer. 
The advantage of this approach is that 
both the electronic and nuclear motion are treated simultaneously 
and it is possible to track and visualize coupled electron-ion dynamics.

The results show that our method can capture the ICD mechanism 
in the hydrogen-bonded water dimer.
By simulating the dynamic process for the hole 
in the hydrogen-bonding proton donor, 
we observed proton transfer occurring with a probability of 20\%, 
leading to the formation of an H$_3$O$^+$ cation and an OH radical fragment, 
which aligns well with existing experimental \cite{18Richter} 
and theoretical \cite{24Kumar} results. 
However, unlike the direct transfer process reported by the BOMD method \cite{24Kumar},
we identified a novel dynamic process in which the proton undergoes a back-and-forth motion before departing again.

When the hole is localized on the hydrogen-bonding proton donor, 
proton transfer is not observed in approximately 80\% of the trajectories. 
In these cases, the proton undergoes a transient back-and-forth motion
but ultimately becomes trapped near water1, with no further transfer occurring. Additionally, when the vacancy is located on the hydrogen-bonding proton acceptor, no proton transfer is observed throughout the simulation.
These findings indicate that, under both conditions, ICD serves as the sole relaxation mechanism, ultimately leading to the formation of two H$_2$O$^+$ fragments. These observations are consistent with the experimental results of Jahnke et al. \cite{10Jahnke}, as well as with the theoretical simulations reported by Kumar et al. \cite{24Kumar} and Wang et al. \cite{24Wang}.

In water-rich environments such as living tissues and other
biological systems, radiation-induced damage is predominantly
observed through ICD. The presented methodology offers an effective 
tool for investigating competing electron relaxation pathways,
with sufficient efficiency to analyze these complex systems. 

\section{Acknowledgements}

\begin{acknowledgments} 
This work was supported by the Natural Science Foundation
of Henan Province under Grant No. 252300421490, and by the National Science Foundation (NSF) 
under Grant No. DMR-2217759. Computational resources were provided by ACES at 
Texas A$\&$M University through allocation PHYS240167 from the 
Advanced Cyberinfrastructure Coordination Ecosystem: Services $\&$ Support (ACCESS) program, 
supported by NSF grants 2138259, 2138286, 2138307, 2137603, and 2138296.
\end{acknowledgments}

\section*{Data Availability Statement}
The data that support the findings of this study are available
from the corresponding author upon reasonable request.

\section*{AUTHOR DECLARATIONS}
\par\noindent
{\bf Conflict of Interest}

The authors have no conflict of interest to disclose.


%

\end{document}